# Insights On Streamflow Predictability Across Scales Using Horizontal Visibility Graph Based Networks

**Ganesh R. Ghimire[1], Navid Jadidoleslam[1], Witold F. Krajewski[1], and Anastasios A. Tsonis[2,3]**

[1] IIHR-Hydroscience and Engineering, University of Iowa, Civil and Environmental Engineering, Iowa City, IA, U.S.

[2] Department of Mathematical Sciences, University of Wisconsin-Milwaukee, Milwaukee, WI, U.S.

[3] Hydrologic Research Center, San Diego, CA, U.S.

[*]**Correspondence:**
Corresponding Author
ganesh-ghimire@uiowa.edu.

**Keywords:** Streamflow Dynamics, Complex Network, Visibility Graph, HVG, Degree Distribution

**Abstract**

Streamflow is a dynamical process that integrates water movement in space and time within basin boundaries. The authors characterize the dynamics associated with streamflow time series data from about seventy-one U.S. Geological Survey (USGS) stream-gauge stations in the state of Iowa. They employ a novel approach called visibility graph (VG). It uses the concept of mapping time series into complex networks to investigate the time evolutionary behavior of dynamical system. The authors focus on a simple variant of VG algorithm called horizontal visibility graph (HVG). The tracking of dynamics and hence, the predictability of streamflow processes, are carried out by extracting two key pieces of information called characteristic exponent, $\lambda$ of degree distribution and global clustering coefficient, $GC$ pertaining to HVG derived network. The authors use these two measures to identify whether streamflow process has its origin in random or chaotic processes. They show that the characterization of streamflow dynamics is sensitive to data attributes. Through a systematic and comprehensive analysis, the authors illustrate that streamflow dynamics characterization is sensitive to the normalization, and the time-scale of streamflow time-series. At daily scale, streamflow at all stations used in the analysis, reveals randomness with strong spatial scale (basin size) dependence. This has implications for predictability of streamflow and floods. The authors demonstrate that dynamics transition through potentially chaotic to randomly correlated process as the averaging time-scale increases. Finally, the temporal trends of $\lambda$ and $GC$ are statistically significant at about 40% of the total number of stations analyzed. Attributing this trend to factors such as changing climate or land use requires further research.

# 1 Introduction

Transport of water in natural streams is one of the main components of the hydrologic cycle. Like other natural systems, streamflow shows fluctuation over time. Intense rainfall events manifesting as peak streamflow and sometimes flooding, longer dry periods followed by low flows and droughts, snowmelt in higher latitudes after cessation of cold season are some examples illustrating streamflow fluctuations. Under changing climate, understanding increased demand of water or flooding issues requires the understanding of variability and predictability of underlying streamflow dynamics. The changing context of human and climate-induced changes in hydrologic cycle manifest into a new realm of streamflow predictability (Kumar, 2011) in addition to its traditional understanding in the context of water management and forecasting. In this study, the authors treat streamflow time-series as an output of the non-linear dynamical system and map it into complex networks using visibility-graph-based algorithm (e.g., Braga et al., 2016; Lacasa et al., 2012; Lacasa and Toral, 2010; Lacasa et al., 2008; Stephen et al., 2015).

Streamflow time series have been studied using different approaches, including Fourier transforms (e.g., Lundquist & Cayan, 2002), wavelet transforms (e.g., Coulibaly and Burn, 2004; Smith et al., 1998), chaos theory (e.g., Bordignon and Lisi, 2000; Porporato and Ridolfi, 1997) and stochastic modeling (e.g., Livina et al., 2003; Prairie et al., 2006; Wang, 2006). Most of these studies were motivated by the needs of streamflow forecasting but determining predictability lies on the ability to distinguish origin of underlying process (e.g. Lacasa and Toral, 2010). The distinction of the presence of low-dimensional chaos (determinism) or long-range randomness (stochasticity) in streamflow has long been a long standing issues that has not been comprehensively resolved.

The issue of mathematical and numerical modeling of river flow time-series to understand the predictability of streamflows requires distinguishing whether the underlying dynamics is deterministic or stochastic. In this direction, limited efforts have been documented in the hydrologic community. Porporato and Ridolfi (1997) and Bordignon and Lisi (2000) checked for the evidence of chaotic behavior considering non-linear dynamics measures such as phase-portrait of the attractor, largest Lyapunov exponent, correlation dimension and demonstrated that presence of a low-dimension chaotic (deterministic) component cannot be excluded. Bordignon and Lisi (2000) further verified the presence of chaotic behavior using the Deterministic versus Stochastic (DVS) algorithm proposed by Casdagli and Weigend (1993). Their study showed that non-linear river flow modeling enhances the predictability of streamflow. Other non-linear method i.e. Correlation Integral Analysis (CIA) was proposed by Grassberger and Procaccia (1983). Pasternack (1999) investigated the presence of low-dimensional chaos using CIA and showed that it is not possible to confirm its presence in streamflow time-series. However, the method could be a useful tool for evaluating model output characteristics. There has not been a clear consensus in the literature regarding identification of chaotic behavior in streamflow time-series in terms of the methods applied, data size used and other factors used in their analyses (see e.g., Pasternack (1999), Koutsoyiannis (2006)).

To shed some more light into this debate, here we apply a new method, which provides a criterion that discriminates between chaotic and stochastic processes. In this method, any time series can be mapped into a network by using visibility graphs (e.g., Braga et al., 2016; Lacasa et al., 2012; Lacasa and Toral, 2010; Lacasa et al., 2008; Stephen et al., 2015). Such graphs have the ability to reveal many salient characteristics of the time series. Regardless of the type of the time series used, the constructed network, by definition, is connected, undirected, and invariant to affine transformations of the time series. Hence, the constructed network from a time series holds its inherent properties.

A simple variant to visibility graph is the horizontal visibility graph (HVG). Mathematically, let $[x_i, i = 1, 2,…, N]$ be a time-series of N data, where $i$ represents nodes. Two nodes $i$ and $j$ are connected if $[x_j, x_j] > x_k \forall k | (i<k<j)$. In other words, two nodes $i$ and $j$ in the graph are assumed connected, if one can draw a horizontal line in the time series joining $x_i$ and $x_j$, which does not intersect any intermediate data height. A key feature of HVG graphs is that the nearest neighbors are visible to each other. Note that rescaling of horizontal and vertical axes, and horizontal and vertical translations do not affect the result of network obtained from HVG graphs. In Figure 1, we illustrate the concept of transforming time-series to a network using HVG. We use this approach to generate the network from streamflow time series to explore its underlying dynamics.

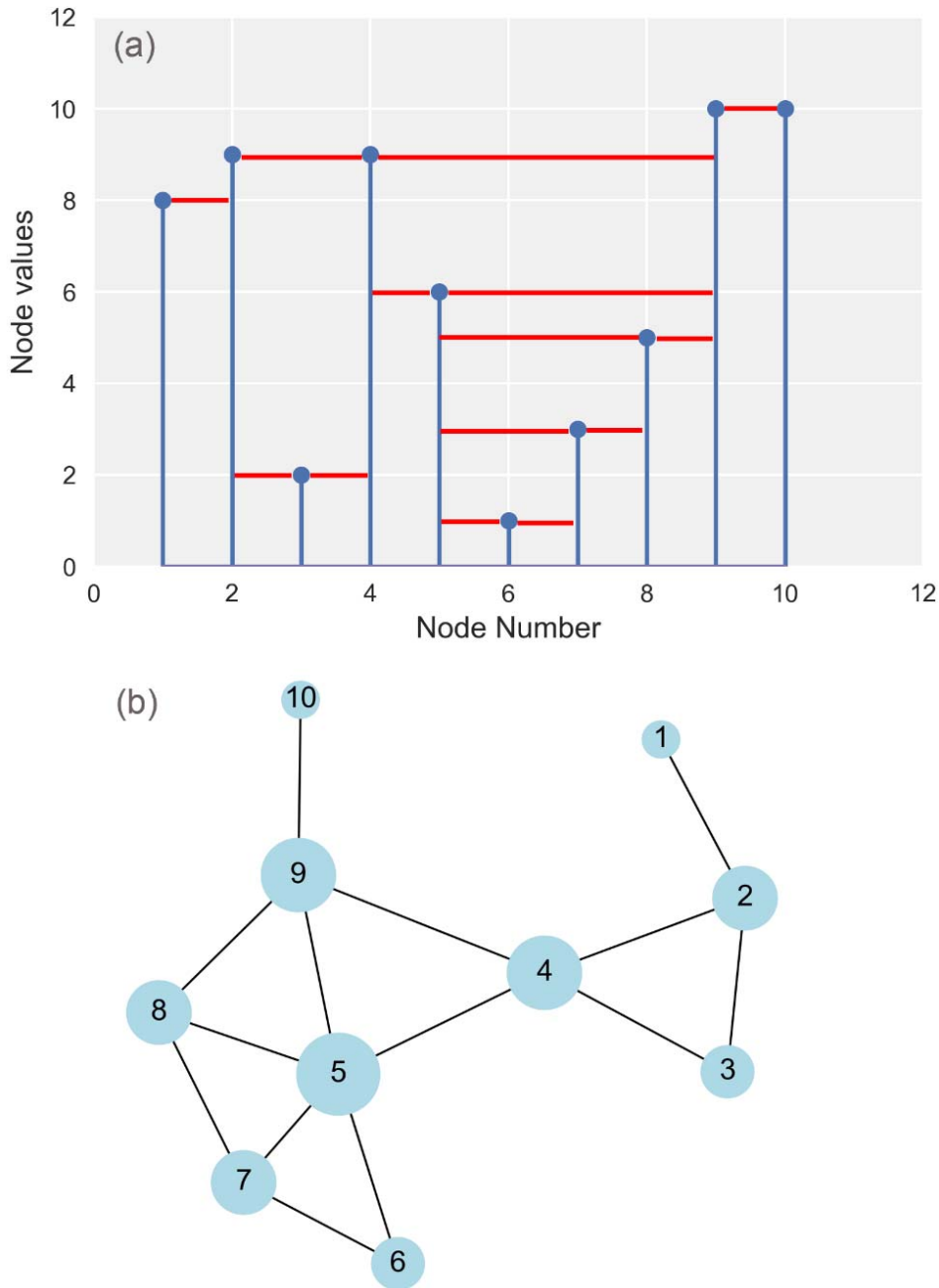

Figure 1: Schematic illustration of horizontal visibility graph (HVG) approach (a) Indices of nodes with corresponding values. The solid red lines show whether corresponding nodes are horizontally visible from each other. For example, take node 4. We can draw a horizontal straight line between node 4 and node 2 that does not intersect the height of node 3. Thus, node 4 is connected to node 2. Similarly, node 4 is connected to nodes 3, 5 and 9, but not, for example to node 6, since there is no way we can draw a horizontal line without crossing the height of node 5. (b) The complete network.

Lacasa and Toral (2010) employed HVG to characterize and distinguish between random and chaotic processes underlying a time series. They showed that time series maps to a network where degree i.e. the number of connections with other nodes have exponential distribution, thus enabling exact distinction between processes. It has been shown analytically (e.g., Braga et al. 2016; Lacasa and Toral 2010) that irrespective of the underlying distributions, the value of characteristic exponent parameter of degree distribution serves as the exact frontier between chaotic and stochastic processes. More recently, Braga et al. (2016) used HVG to analyze river flow fluctuations at daily time-scale using 141 stream gauges in Brazil. They demonstrated the presence of correlated stochastic structure in the streamflow dynamics through degree distribution and global clustering coefficient measures of the resultant network. Further, Lange et al. (2018) investigated the sensitivity of the HVG methodology to streamflow time series pre-processing properties such as time series length, presence of ties and deseasonalization using around 150 time series from regulated rivers in Brazil at daily time-scale. They showed that data pre-processing can result in contradictory results and thus should be used with caution. In another study, Gonçalves et al. (2016) explored new ways to extract information from HVG using information theory. They showed that alternative distributions to degree distributions such as distance distribution and weight distribution can help extract efficient information especially for shorter time-series. Further, Serinaldi and Kilsby (2016) used a directed HVG to explore irreversibility, a signature of nonlinearity of streamflow through analysis of degree distributions. In their comprehensive study, they used 699 unregulated daily time scale streamflow time-series across the conterminous United States (CONUS) to show that degree distributions have systematic sub-exponential behaviors of different strengths and quantified it through the information theory measures. Their findings show that streamflow dynamics are more complex than simple stochastic linear dynamics and irreversibility is a key feature.

In this work, we study unregulated streamflow time series at United States Geological Survey (USGS) stream gauge stations in the State of Iowa in U.S using networks derived from simple undirected HVG. Our objective is not just to explore the predictability of streamflow in terms of processes generating them but also to answer some questions not addressed by previous works. Specifically, we address the following questions: (1) What process underlies streamflow dynamics? (2) How does time resolution of streamflow time series data, as well as a normalization procedure, impact inference from HVG based networks? (3) Does the description of this process demonstrate spatial (basin) scale dependence? (4) Do the characteristics describing this process show temporal evolution?

This paper is organized as follows. In the methods section, we present the study area, data, and provide HVG application strategy across time-scales of streamflow time-series. In the results and discussion section, we seek answer(s) to the above questions based on results from our analysis. Finally, we draw conclusions from this work presenting avenues for further research on visibility-based network analysis of streamflow dynamics in hydrologic context.

## 2 Methods

We conducted this study in the domain of the State of Iowa with rivers draining to the Mississippi and Missouri Rivers. At present, around 140 USGS streamflow gauges monitor the streams and rivers providing data for this study. Figure 2 shows the spatial distribution of these stations over Iowa. The stream gauges measure drainage areas that range from about 7 km$^2$ to 37,000 km$^2$. We considered USGS daily streamflow records with record length of at least 50 years. About seventy-one stations qualify this criterion, as depicted by dark green dots in Figure

2. There are about six stations out of seventy-one which are distant downstream from reservoirs. However, we expect them to have minimal effect on our overall inference from hydrologic standpoint of streamflow predictability. Figure 3(a) illustrates the distribution of basin scales and Figure 3(b) illustrates the distribution of daily streamflow records length of USGS stations, respectively. The wide range of basin scales monitored enables us to capture the spatial scale dependence of HVG derived information measures.

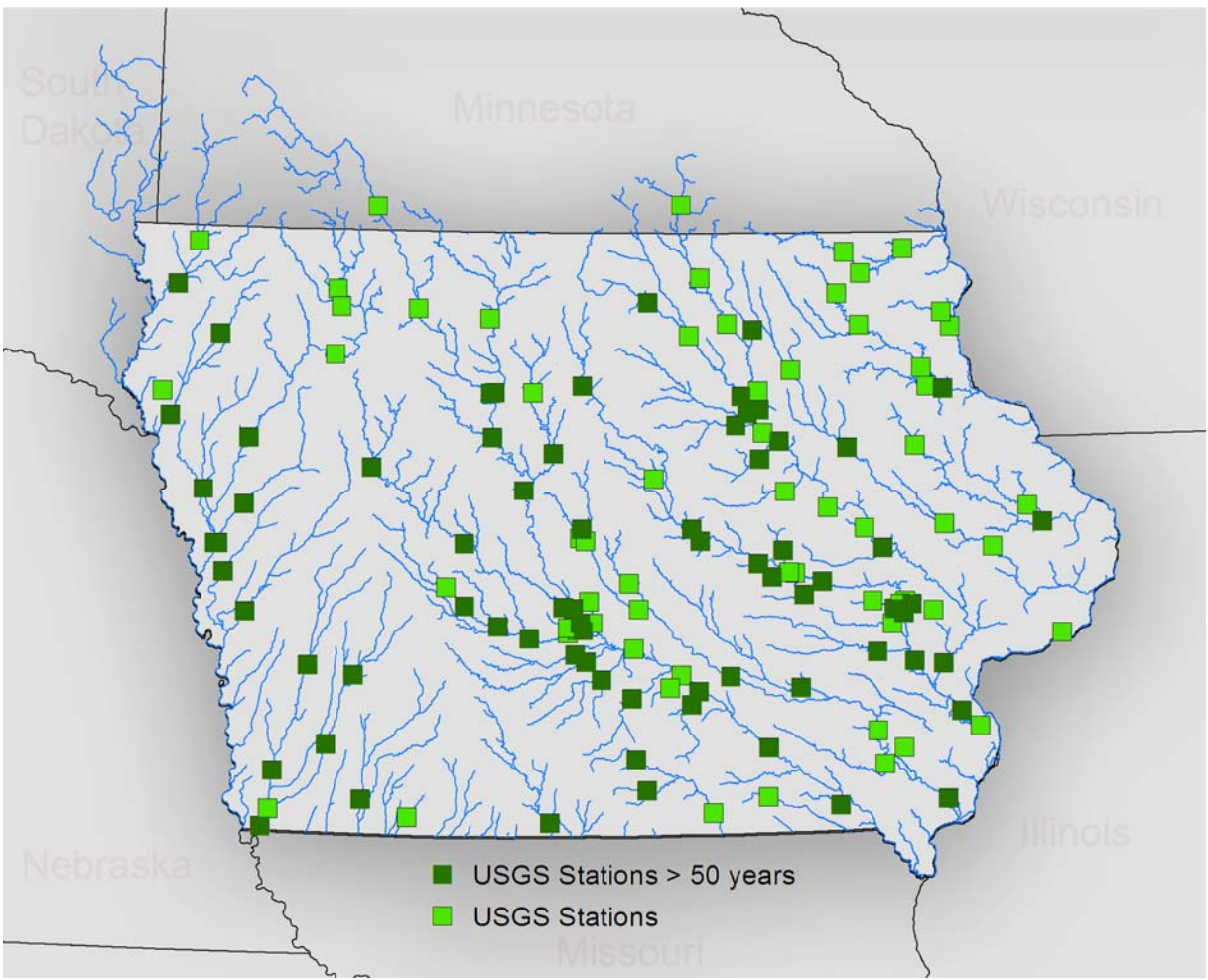

Figure 2: Spatial distribution of USGS stations (both light and green squares) across the State of Iowa. The dark green squares represent USGS stations with at least 50 years of streamflow records.

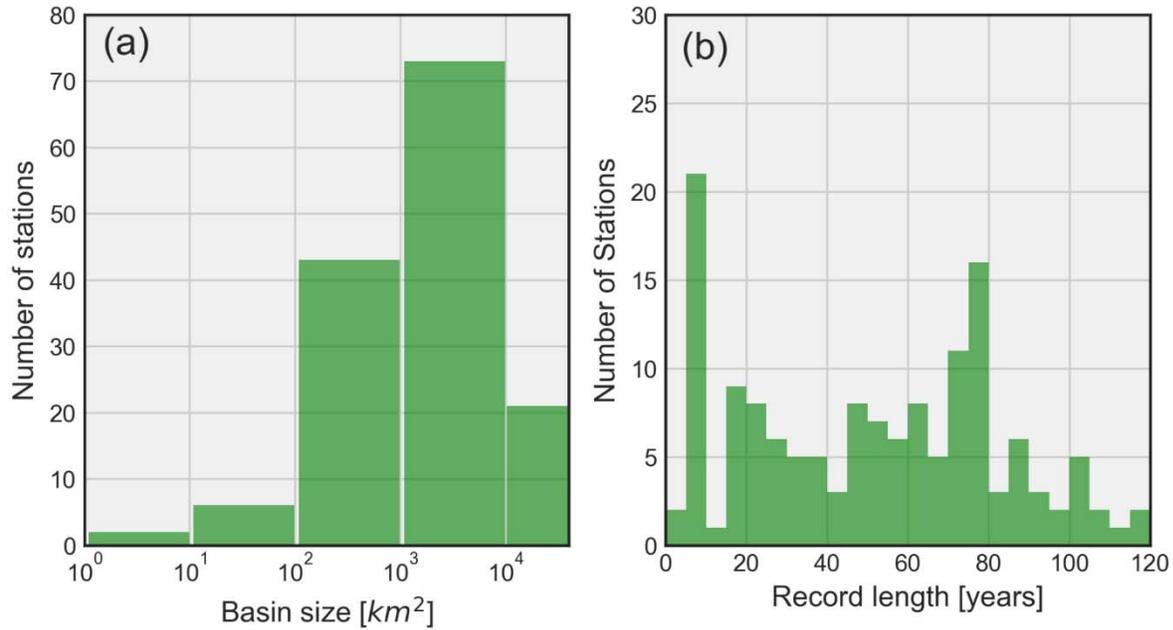

Figure 3: Histogram of USGS basin characteristics (a) distribution of basin scales across the State of Iowa; (b) distribution of streamflow record lengths in terms number of stations.

From river network standpoint, about 65% of the state drains to the Mississippi River while about 35% of the state drains to the Missouri River (e.g., Ghimire et al., 2018; Ghimire et al., 2019). Most of the land use in the state is predominantly agricultural. The North-eastern part of the state can be described by deeply carved terrain, narrow valleys and relatively higher stream slopes while low-reliefs with relatively milder stream slope represent the rest of the state (e.g., Ghimire et al., 2019). As our goals include exploring the impact of time resolution of streamflow time series on our inference, we ought to use station with different record lengths. At daily time-scale, we used data from seventy-one stations with records of 50 years or longer. However, for comparison of three time-scales: instantaneous (15 minutes), hourly, and daily, we use records between 2002 and 2018 as prior to 2002 only daily data are available. We adopted 15-minutes streamflow time series first and averaged them to generate subsequent hourly (four nodes) and daily scale (ninety-six nodes) streamflow time series for the entire state.

We constructed networks representing every year of the historical streamflow records for USGS stations described above. For each year there are 365 nodes represented by the index $i = 1, 2, \ldots, 365$ such that $N = 365$ at daily time-scale. To avoid major seasonal trends between years, we normalized the time series using the mean and standard deviation of flows for each day of the year. We discuss later the impact of this normalization on inference derived from network. Let $X_t(i)$ represent the flow in the year '$t$' on the day '$i$'. Likewise, let $x_t(i)$ be the normalized flow for the year '$t$' on the day '$i$'. Then, we define the normalization of time-series through equations (1) - (3).

$$x_t(i) = \frac{X_t(i) - \mu(i)}{\sigma(i)} \tag{1}$$

where

$$\mu(i) = \frac{1}{n} \sum_{t=1}^{n} X_t(i) \tag{2}$$

and

$$\sigma(i) = \sqrt{\frac{1}{n-1} \sum_{t=1}^{n}(X_t(i) - \mu(i))^2} \tag{3}$$

represent the mean and standard deviation of the flow respectively on the day '$i$' obtained from the distribution over the years of historical records at the given station ($n \in [50, 115]$). As an illustration, we present in Figure 4 envelope of the entire record of daily streamflow together with its mean and median (see Figure 4(a) and 4(c)) as well as the corresponding envelope of the normalized flows (see Figure 4(b) and 4(d)) for a large basin (16,800 km$^2$) of Cedar river at Cedar Rapids and a small basin (70 km$^2$) of Rapid Creek near Iowa City respectively. What is apparent is that the seasonality of flows clearly visible in the original series is much reduced in the normalized data.

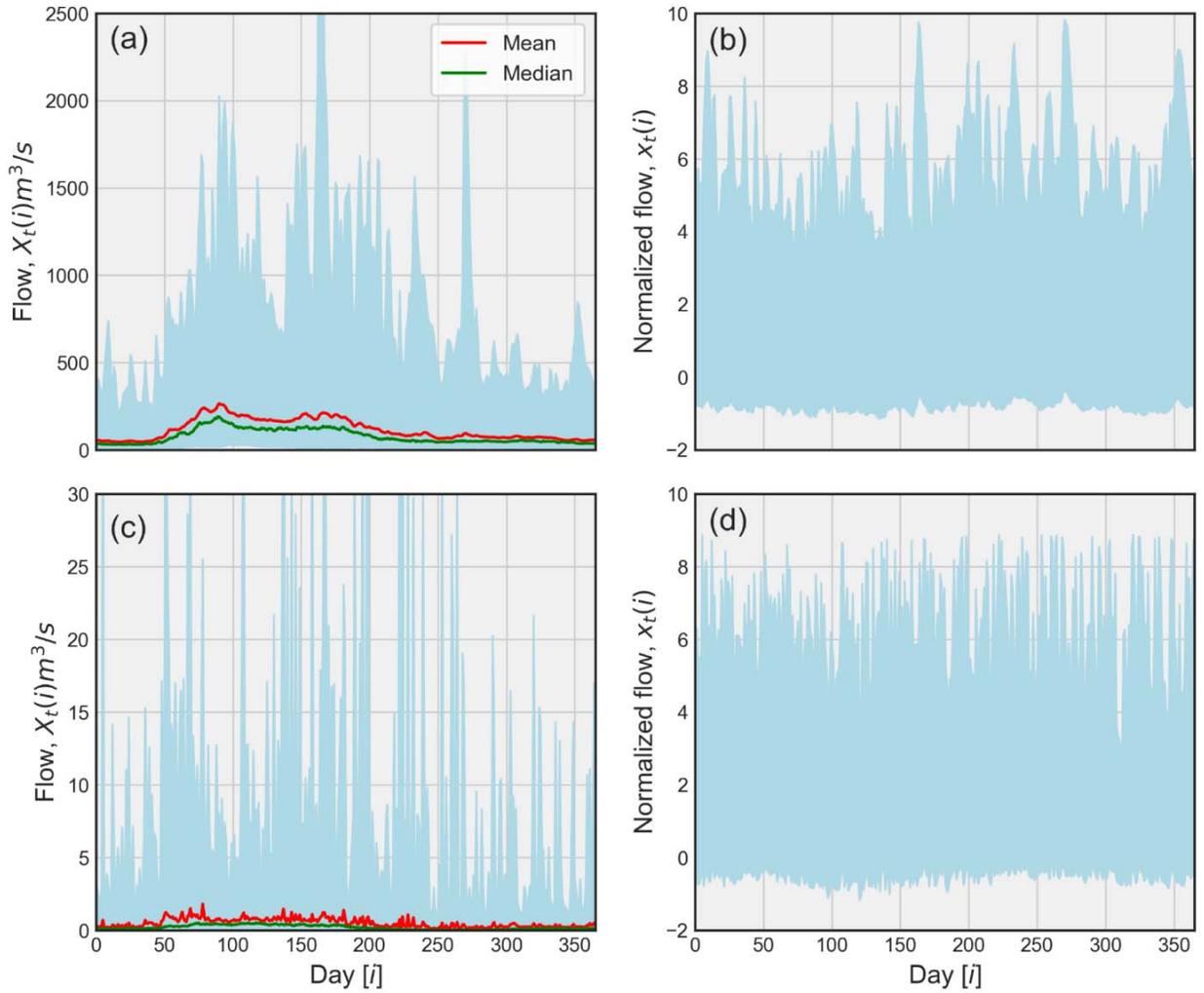

Figure 4: Streamflow time series normalization for two basins. (a) Shaded envelope representing an ensemble of entire raw streamflow records, $X_t(i)$ at daily scale for Cedar River at Cedar Rapids (16,800km$^2$). This envelope shows the variability of streamflow, $\sigma(i)$ while the red and green lines represent mean, $\mu(i)$ and median of flows respectively. (b) Shaded envelope representing an ensemble of normalized streamflows, $x_t(i)$ corresponding to (a). (c) Raw streamflow time series at Rapid Creek near Iowa City (70 km$^2$) with same description as (a). (d) Normalized streamflow time series corresponding to (c).

Next, we mapped the time series for every year using HVG to a complex network. An illustration of the network associated with two streamflow time series is presented in Figure 5.

Figure 5(a) corresponds to the network associated with normalized time-series at Cedar River at Cedar Rapids (16,800 km$^2$) while Figure 5(b) shows the network of streamflow obtained from the same daily data but shuffled randomly over time. The color code represents the month of a year associated with each node with their size representing the number of connections they make with their neighbors. Clearly, larger nodes of HVG derived network are associated with the major events. As Figures 5(a) shows, the nodes related to consecutive months (neighbors in the proximity) are connected to each other implying the information transfer during streamflow generation process in the form of antecedent soil moisture, baseflow, and others.

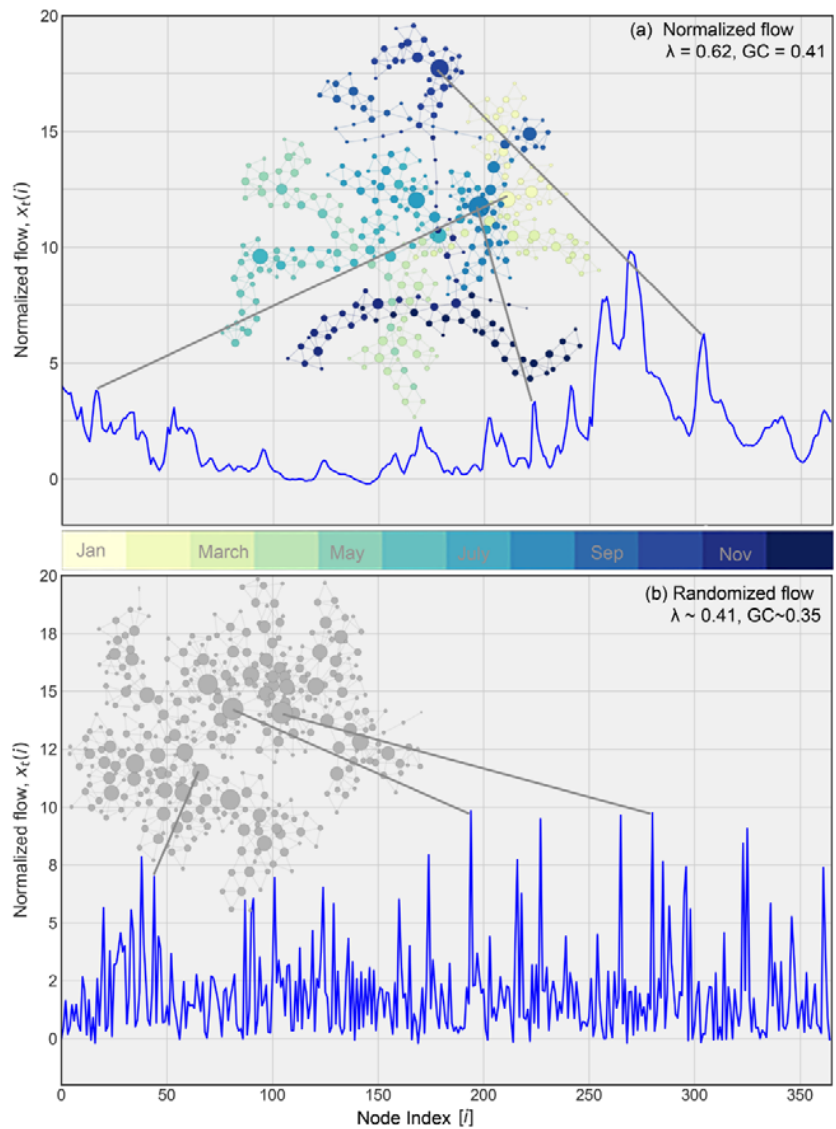

Figure 5: HVG derived network for Cedar River at Cedar Rapids (for the year 2016). (a) Network associated with normalized flow (corresponds to Figure 4). Each color represents the corresponding month of a year where size of each node corresponds to the number of connections with its neighbors. (b) Network for randomly shuffled streamflow from (a). In the normalized data, λ>0.41 indicating the presence of a stochastic process. Both λ and *GC* are lower in the shuffled time series, as expected.

We extracted two fundamental pieces of information from these networks to explain the underlying streamflow dynamics and hence, the streamflow predictability. First metric is the degree distribution. The degree, *k* corresponds to the number of connections a node can have with other nodes. An ensemble of nodes results in a distribution of *k* with *P(k)* representing its cumulative probability function i.e. *P(K ≥ k)*. There are documented efforts (e.g., Braga et al., 2016; Lacasa et al., 2012; Luque et al., 2009) illustrating that *k* resulting from HVG follows the exponential distribution of the form,

$$P(k) \sim e^{-\lambda k} \qquad (4)$$

where $\lambda$ is the decay parameter, also referred to as the characteristic exponent. For the purely random process, it has been analytically demonstrated (e.g., Braga et al. 2016; Lacasa and Toral 2010; Lacasa et al., 2012; Luque et al., 2009) that $\lambda = \lambda_{rand} = ln\left(\frac{3}{2}\right)$. For logistic map, an example of a deterministic chaotic time series, $\lambda$ from HVG is equal 0.26 (see Lacasa and Toral, 2010). The decay parameter for the purely random process serves as the limit for describing underlying behavior of the streamflow process. If $\lambda < \lambda_{rand}$, the streamflow process is chaotic i.e. the dimensionality of the system is smaller. If $\lambda > \lambda_{rand}$, the streamflow is a stochastic process with some dependence structure, e.g. linear. In such a case, the process will have higher predictability that can lead to higher forecasting skill. For our analysis, we obtained the degree distribution for every year at each station.

The second information we extracted from the HVG derived network is called the global clustering coefficient, $GC$. It is a measure of likelihood of nodes to form clusters of tightly-knit groups. Because we explored only the clustering nature of the entire network rather than its local behavior, only the computation procedure of global clustering coefficient is discussed here. We computed it as the ratio of number of closed triplets (or three times number of triangles) to the total number of triplets (both open and closed) in a network. For example, the global clustering coefficient for the network in Figure 1 is equal to 15/33=0.45. The five triangles in center show the clustering of these nodes dominating the entire network. The range of values of $GC$ is [0-1] with 1 corresponding to a full network of triangles. As $GC$ approaches the value of 1, the network becomes more and more fully connected, which means that every node is connected to all other nodes, which in turn means that every node "sees" every other node, hence the process is perfectly linear.

To summarize: (a) If $\lambda$>0.41, then the process is a red noise (stochastic) process and thus linear. (b) If $\lambda$<0.41, then the process is a chaotic (nonlinear) process. $GC$ is expected to be higher in the former cases than in the later cases. It follows that the estimation of $\lambda$ and $GC$ provides insights on the predictability of the process in questions since a chaotic process is inherently more unpredictable that a linear stochastic process.

In the context of normalized flow of Cedar River at Cedar Rapids (see Figure 5(a)), the major events are less dominant in the entire network resulting in relatively simpler internal network structures. Consequently, the values of $\lambda$ and $GC$ are higher, and hence the higher streamflow predictability. The corresponding randomly shuffled time series in Figure 5(b), however, shows event-like signals (not true streamflow signal) dominating the entire network resulting in a random internal network structure. The resultant values of $\lambda$ and $GC$ are as expected smaller. In this study we computed $GC$ for every year at each station in the same way as we did for the degree distribution.

## 3 Results and Discussion

### 3.1 Normalized Streamflow Time Series

In Figure 6, we present degree distributions for the Cedar River at Cedar Rapids along with procedure for the computation of $\lambda$ as a demonstration of the process using daily values. The results shown here follow from Figures (4) – (5). We fit a linear regression model to degree distribution for each year in log-linear space (see Eq. 4) such that the slope parameter of the model fit corresponds to $\lambda$. We exclude non-exponential part of the degree distribution and considered degree, $k > 2$ for the fit. Virtually all gray lines in Figure 6, which represent degree distribution for each year, show $\lambda$ greater than $\lambda_{rand}$. It is apparent from the figure that the uncertainty increases with the increase in $k$. The larger the values of $k$, the larger the number of connections in the network corresponding to the ability of streamflow peaks to see through horizontally a greater number of neighbors. As illustrated by the dark solid line, average $\lambda$, $\langle \lambda \rangle > \lambda_{rand}$ suggesting that the process underlying streamflow dynamics is a correlated stochastic process. In other words, larger likelihood of smaller degrees signify that nodes of time-series have longer correlations. The presence of correlated structure in streamflow dynamics indicates strong predictability of streamflow.

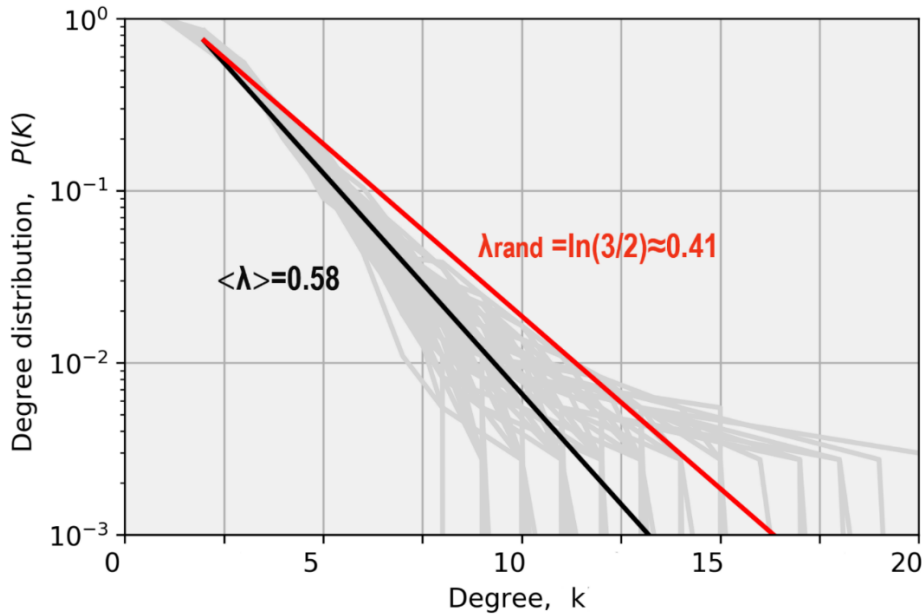

Figure 6: Illustration of degree distributions for the Cedar River at Cedar Rapids. Grey lines represent degree distributions for each year (daily values), while solid black line represents the mean exponential fit. The solid red line corresponds to the degree distribution associated for a purely random process. Note that, basically for all years, λ is greater than 0.41, indicating a consistent stochastic process.

To elucidate this further, consider the violin plots in Figure 7, which is an improved version of box plots with kernel density smoothing. This plot shows the variability of *GC* in the moving window of *λ* capturing the overall strong relationship between λ and *GC*. We show here values of *λ* and *GC* computed across all stations for all years pooled together. It is apparent that all stations exhibit $\lambda > \lambda_{rand}$. Virtually all values of *GC* > 0.345 (see Braga et al., 2016) suggesting that the streamflow process for the entire Iowa region at daily time-scale has its origin in the long-range correlated stochastic process. In other words, both measures can aptly reveal the stochastic nature of the underlying streamflow process and the associated potential predictability at daily time scale. Because this plot shows overall dynamics of the process, it might not sufficiently illustrate the temporal evolution of each of these measures. Therefore, we discuss the evolutive nature of $\lambda$ and $GC$ in more detail in the next section.

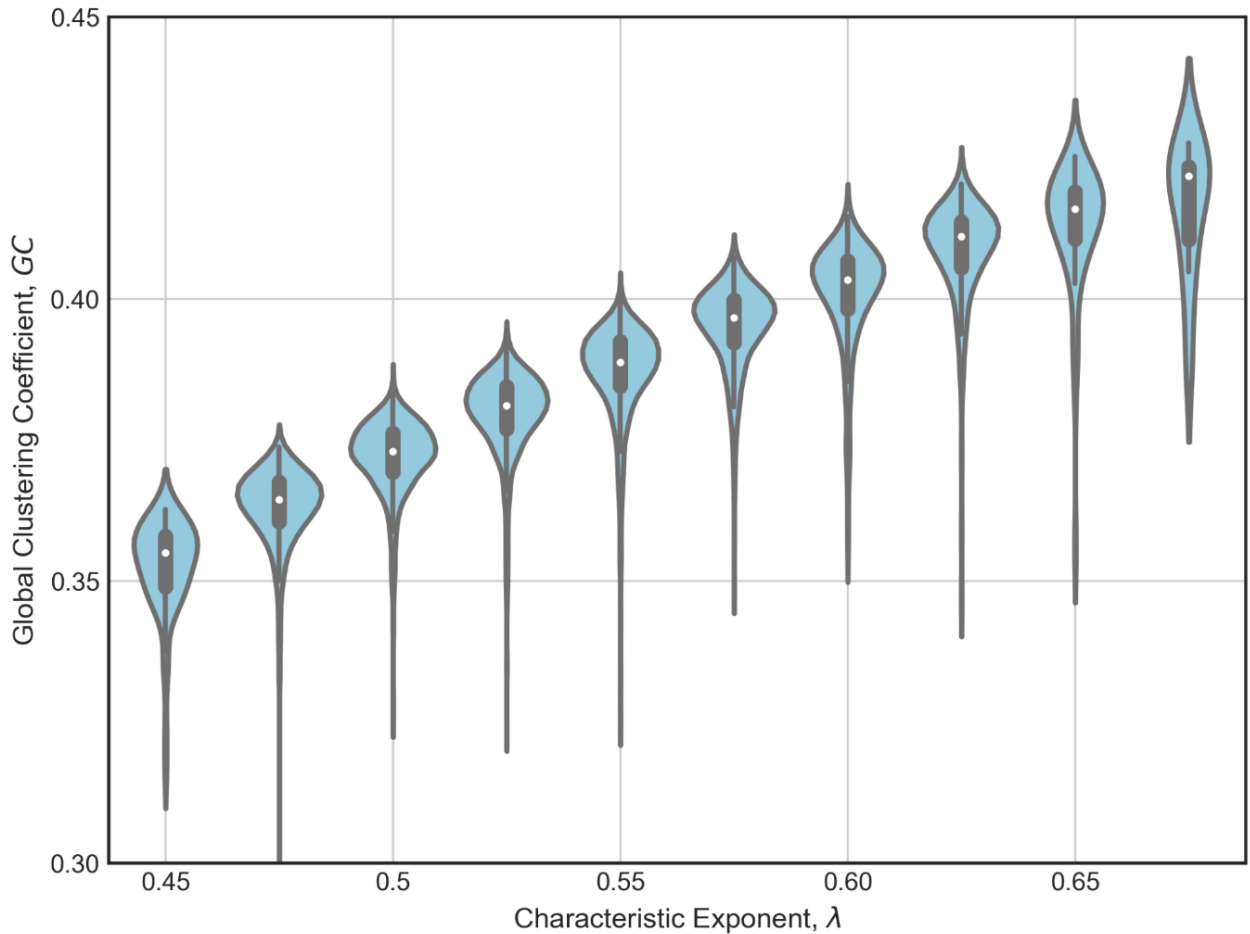

Figure 7: Plot showing a strong relationship between characteristic exponent, $\lambda$ and global clustering coefficient, $GC$ for all stations and all years pooled together. The violin plots show the distribution of $GC$ for the moving window of $\lambda$. Each violin plot is an extended form of boxplot showing the median in white circle with solid dark line depicting interquartile range while shape of the violin shows the point density distribution with kernel smoothing. It is apparent that all stations exhibit $\lambda > \lambda_{rand} = 0.41$. Virtually all values of $GC > 0.345$ (Braga et al., 2016) suggesting that the streamflow process for the entire Iowa region at daily time-scale has its origin in the long-range correlated (persistence) stochastic process. Note that as $\lambda$ increases, $GC$ increases too, as expected when the process becomes more and more linear.

### 3.2 Natural (Raw) Streamflow Time Series

A rationale behind using normalized streamflow time series is to avoid potential seasonal trends in streamflow. Here, we perform similar analysis for $\lambda$ and $GC$ using natural and raw streamflow time series. In Figure 8, we present two-dimensional histograms with 1:1 relationship between metrics for raw and normalized streamflow for the pooled data for the state. The distribution of $\lambda$ around 1:1 line is almost symmetric as depicted by the percentage of pooled data. It illustrates that $\lambda$ derived from HVG based network (see Figure 8(a)) is not overly sensitive (47% versus 53% toward normalized and raw data respectively) to the normalization of streamflow, at least at the daily time-scale. In Figure 8(b), we show the histogram for $GC$ between normalized and raw streamflow data. Unlike $\lambda$, $GC$ shows strong bias toward the normalized data (99%). As can be seen from the figure, the dynamic range of $GC$ is much smaller compared to $\lambda$. The disparity in our inference from $GC$ arises mainly owing to the disparity in complex internal structures of the network for two forms of streamflow time-series (see Figure 9). Figure 9 shows an example of the difference in network structure for Cedar River at Cedar Rapids for years 1904 and 2016. The natural streamflow data for the year 1904 (see Figure 9(b)) shows significant departure in the network internal structure from the year 2016 (see Figure 9(a)) mainly due to dominant base flow conditions in the network. Consequently, $\lambda$ is enhanced while $GC$ shows decline because the internal structure of the network is simplified. This is reflected in a large majority of $GC$ obtained from natural streamflow depicting chaotic regime. It is at least illustrative of the fact that $GC$ is more sensitive to the form of streamflow data in assessing the streamflow predictability based on HVG derived network.

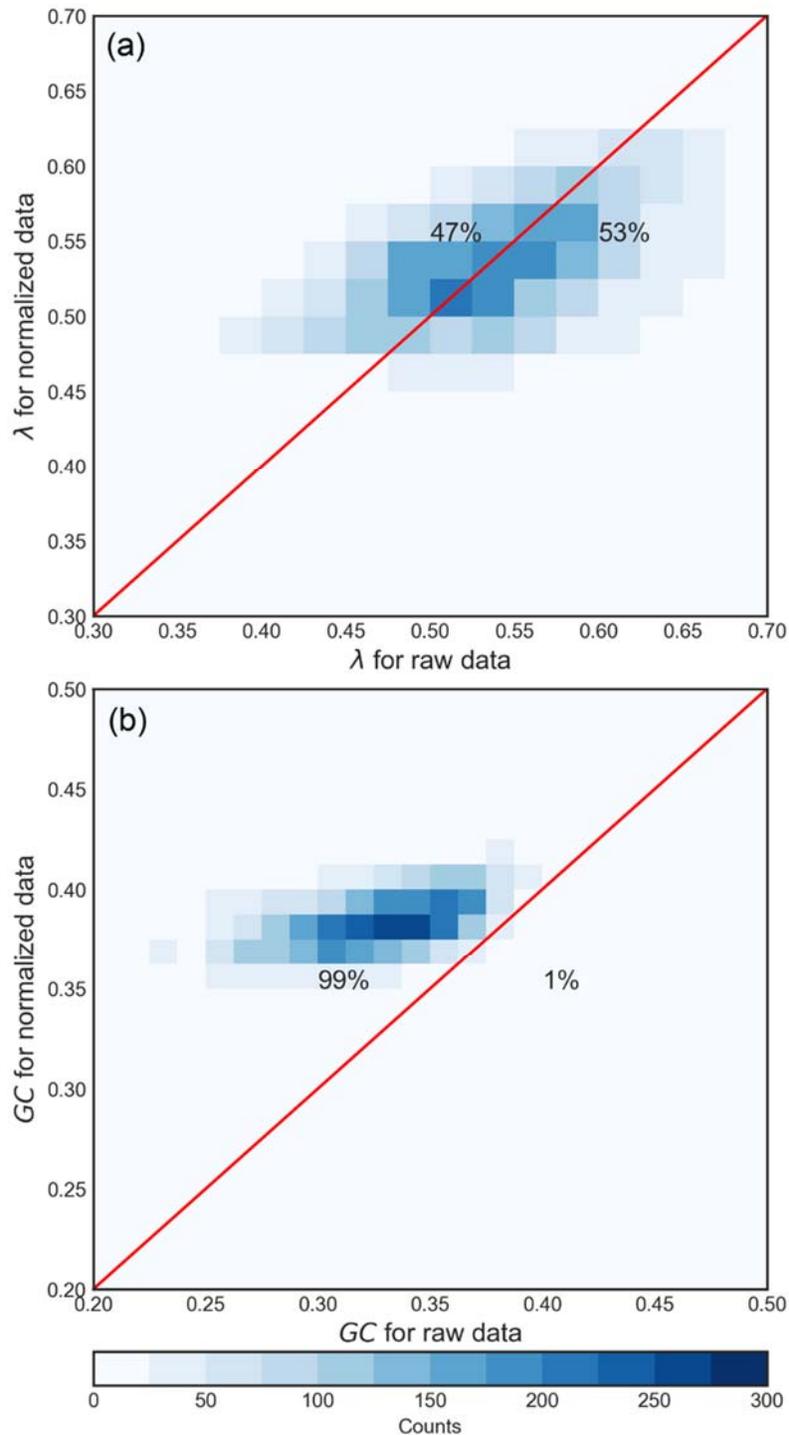

Figure 8: Two-dimensional histogram depicting the comparison between metrics for normalized and raw streamflow time series. The metrics are pooled from all stations for all historical records. (a) Comparison of $\lambda$, and (b) Comparison of $GC$. As it is explained in the text, $GC$ is more affected by normalization than $\lambda$, with higher values for the normalized data, indicating that normalization tends to result in more connected networks, hence more linear processes.

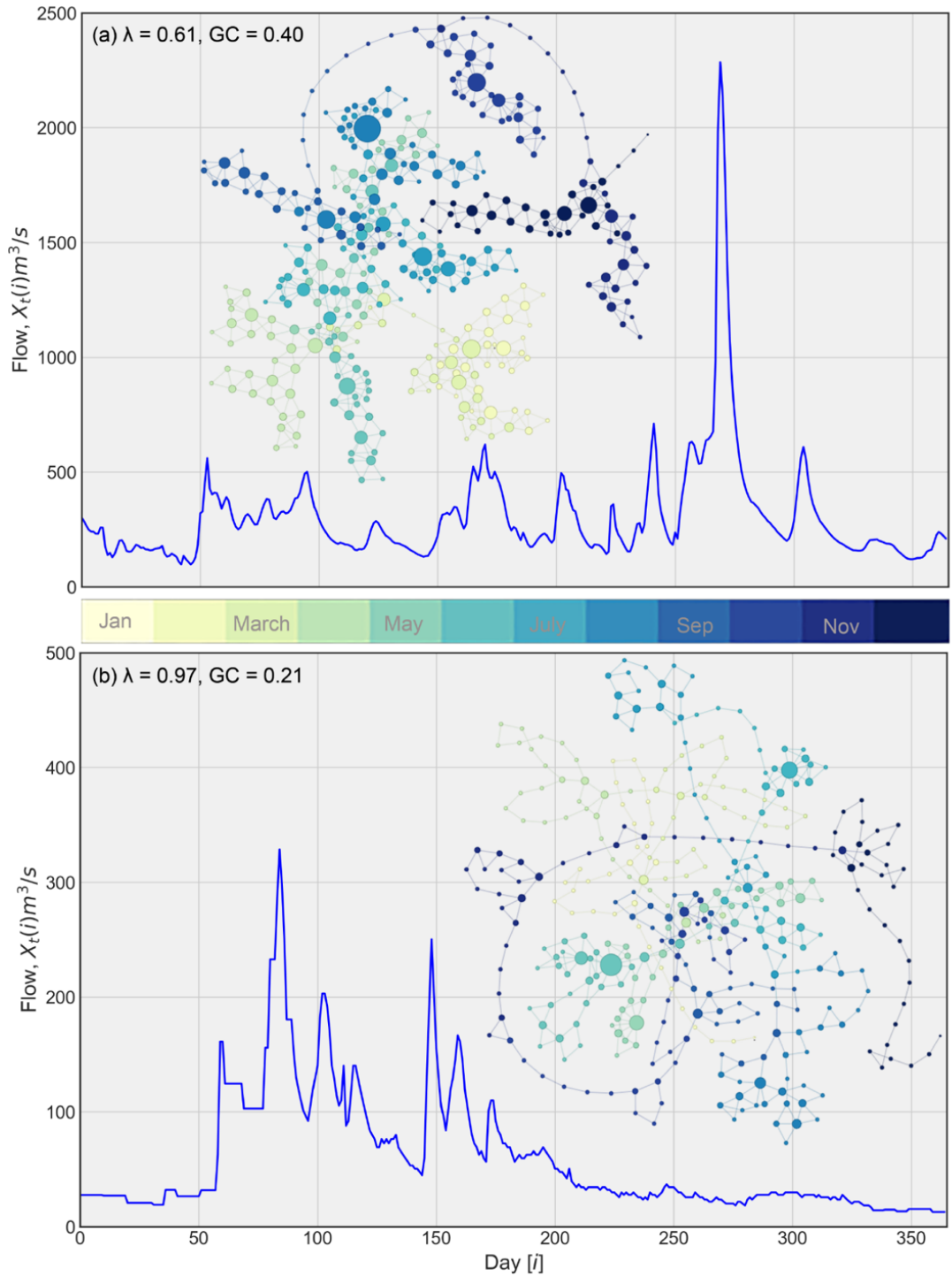

Figure 9: Illustration of networks corresponding to natural streamflow time series for Cedar River at Cedar Rapids station. The color code corresponds to the month of the time series while the size of nodes corresponds to the number of connections they have with other nodes. (a) Network for the year 2016. (b) Network for the year 1904.

### 3.3 Effect of Time Scale of Streamflow Time Series

With the availability of instantaneous (sampled every 15-minutes) streamflow data, we want to exploit its utility especially in the context of streamflow forecasting. Therefore, we devise an experiment to explore the variability of the predictability measure $\lambda$ across three timescales *viz.* 15-minutes, hourly, and daily using the same set-up as described earlier for the daily time-scale streamflow time series with entire historical records. Note that we use the normalized streamflow for all three scales.

In Figure 10(a), we present the variability of $\lambda$ across the three time-scales. It demonstrates that variability of $\lambda$ in terms of interquartile range (thicker solid line inside each violin) do not differ much across time-scales. However, the major difference is in terms of the median values of $\lambda$. Clearly, the mean and median values of $\lambda$ show the decrease as we transition to finer time-scale. The mean and median $\lambda$ falls marginally below $\lambda_{rand}$ limit showing that there is a tendency to transition to chaotic dynamics from stochastic dynamics as we move to finer resolution. In other words, the predictability of streamflow process is sensitive to time-scale of streamflow time-series. Figure 10(b) shows similar behavior across time-scales for *GC*. The mean and median *GC* show increase as we increase the time-scale. The *GC*, however, interestingly shows the reduction in the variability as the averaging time-scale increases, which was not as apparent with $\lambda$.

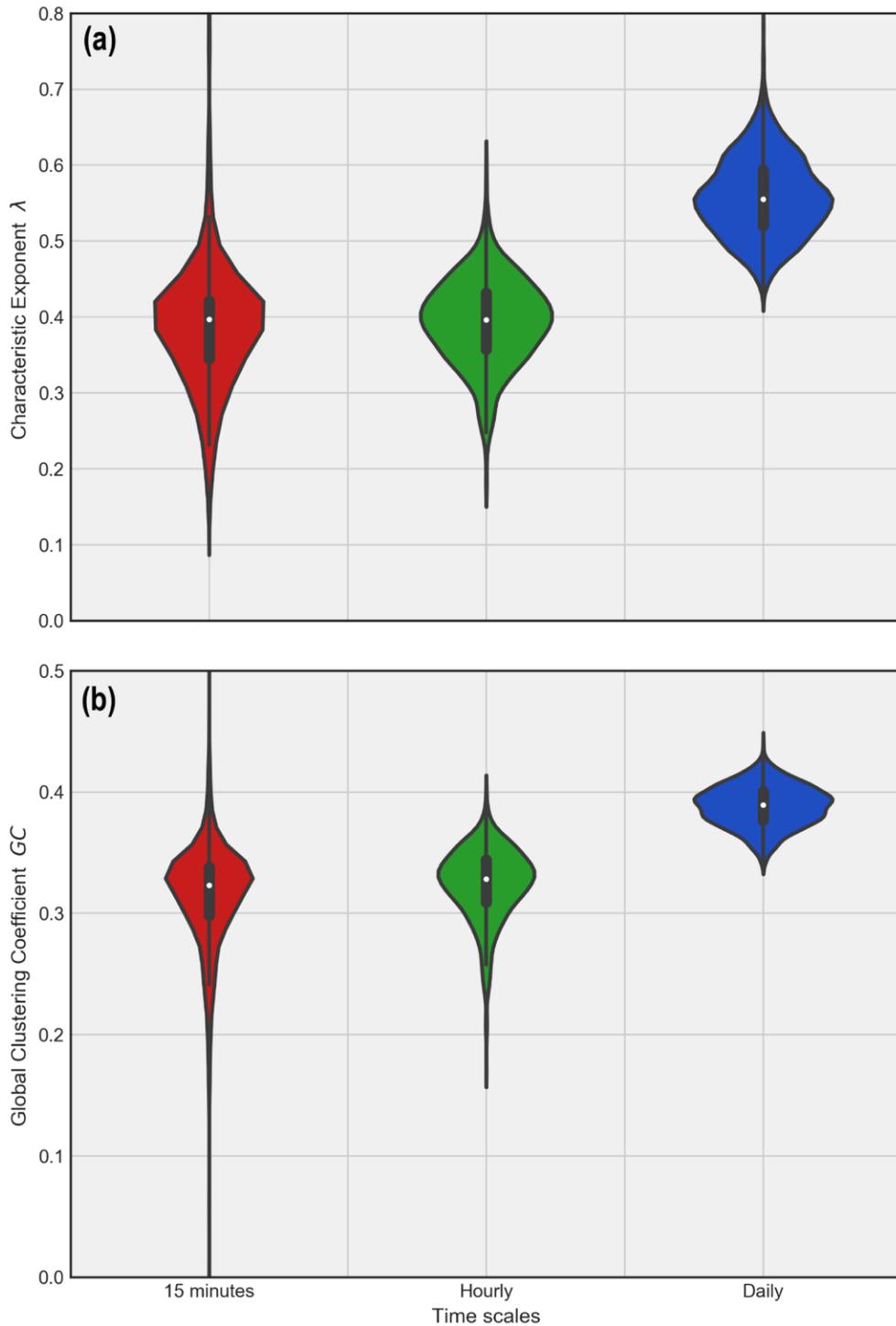

Figure 10: Violin plots of $\lambda$ and $GC$ across time-scales of streamflow time series in (a) and (b) respectively. The solid white dot inside each violin represents median while the thicker solid line inside represents interquartile range. Clearly, the daily time-scales have higher $\lambda$ and higher $GC$ indicating that averaging affects the dynamics (from chaotic at short time-scales to stochastic at longer time-scales).

### 3.4 Spatial and Temporal Dependence of $\lambda$

From hydrologic standpoint, it is important to understand the predictability of streamflow process across spatial scales. In Figure 11, we illustrate the relationship between $\lambda$ and basin scale. Figure 11(a) depicts a clear spatial dependence of $\langle \lambda \rangle$ emerging from the daily scale streamflow time-series. The solid black line represents the predictive model fit using power law which shows that the variability in streamflow predictability in terms of $\lambda$ is explained by the basin scales. For detailed visualization of uncertainty in $\lambda$ conditional on basin scales, we present violin plots in Figure 11(b). The uncertainty arises from variability over the historical periods of records conditional on windows of basin scales. Most of the point density of $\lambda$ is around the median depicting the similar relationship as Figure 11(a). Clearly, larger the basin scales, longer the persistent behavior of streamflow (Ghimire and Krajewski, 2019) and hence the predictability of streamflow. This is supported by the fact that larger basin scales have been shown to display long-memory. This result has important implication for the water resources management and skillful streamflow forecasting.

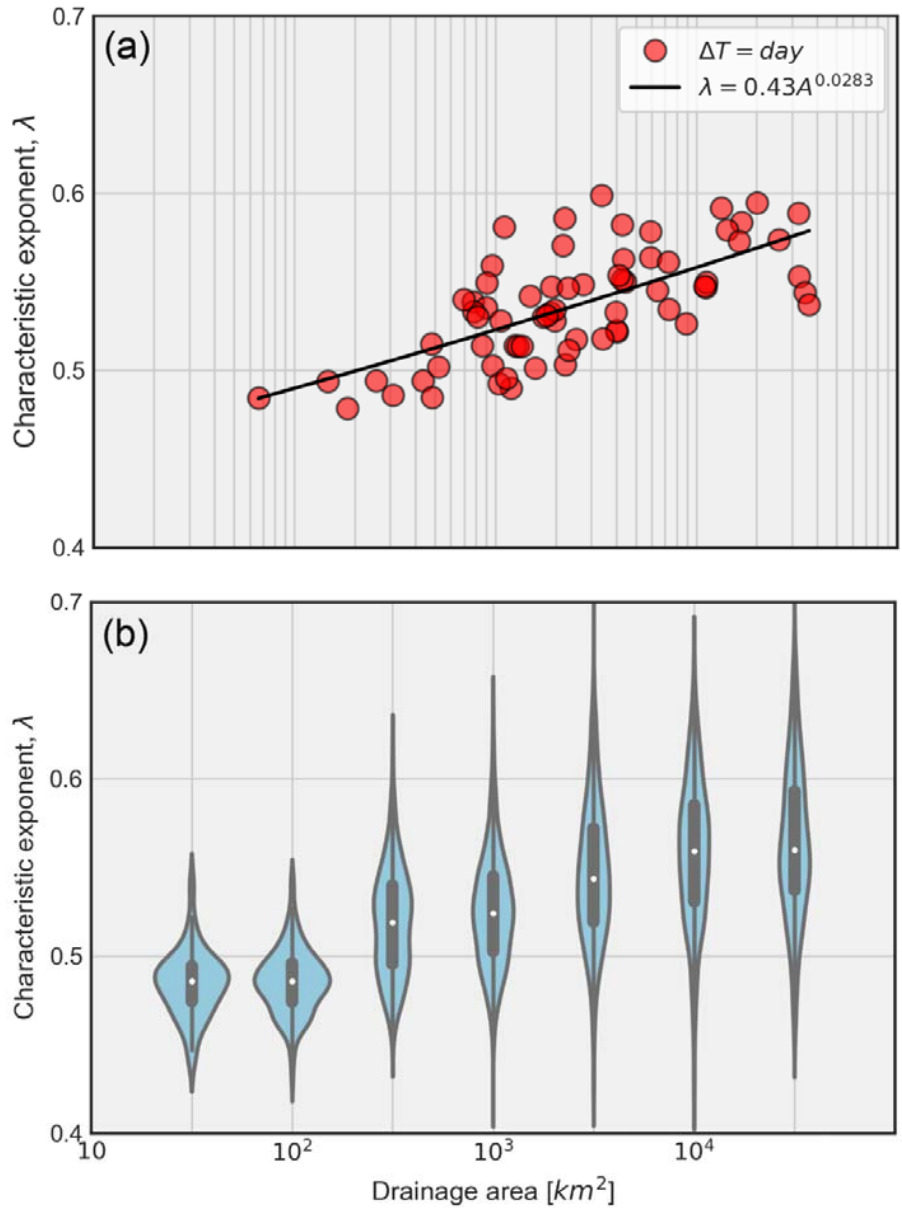

Figure 11: Spatial dependence of $\lambda$. (a) Expected value of $\lambda$ over the years across stations. The solid dark line corresponds to the power-law model fit to the data. (b) Distribution of $\lambda$ over the years considering pooled data from consecutive windows of basin scales. Clearly, the size of drainage area affects the resulted processes. The larger the basin scales, the longer the persistent behavior of streamflow.

In addition, it is of interest to forecasting community to assess the evolutive trend of streamflow predictability measures $\lambda$ and $GC$. To explore this trend, we fit a linear regression model using the ordinary least square method for each measure with $t$ [years]. For $\lambda$, the model we fit is of the form:

$$\lambda = a + bt \qquad (5)$$

where $a$ and $b$ are intercept and slope of the model fit line respectively. Subsequently, we fit similar model as Eq. 5 to $GC$ of the form:

$$GC = a^* + b^*t \qquad (6)$$

where $a^*$ and $b^*$ are intercept and slope parameters of linear model fit respectively employing ordinary least squares method. Figure 12 shows spatial distribution of stations with evolutive trend of predictability measures $\lambda$ and $GC$. The circles with partial blue and red colors correspond to the stations with statistically significant values of $b$ and $b^*$ at 95% confidence level. In other words, twenty-two such stations show significant evolutive trend of both $\lambda$ and $GC$. The blue and red circles represent stations with significant evolutive trend of $\lambda$ and $GC$ alone respectively. It shows that both measures depict evolutive trend at virtually similar number of collocated stations across the state (see Figure 12). Though the assumption of residuals distribution being normal is not ideal in this case, studies have shown that using the bootstrap regression also yields similar results (Braga et al., 2016). Moreover, a large majority of these stations are of scales larger than 1000 km$^2$, which suggests that evolutive trend of streamflow predictability is more prominent at larger basin scales. Further, it is apparent at the central-eastern parts of the state which could be attributed to factors such as anthropogenic and climatic changes, and modifications to the base flow conditions. The explicit attribution of trend is beyond the scope of this study.

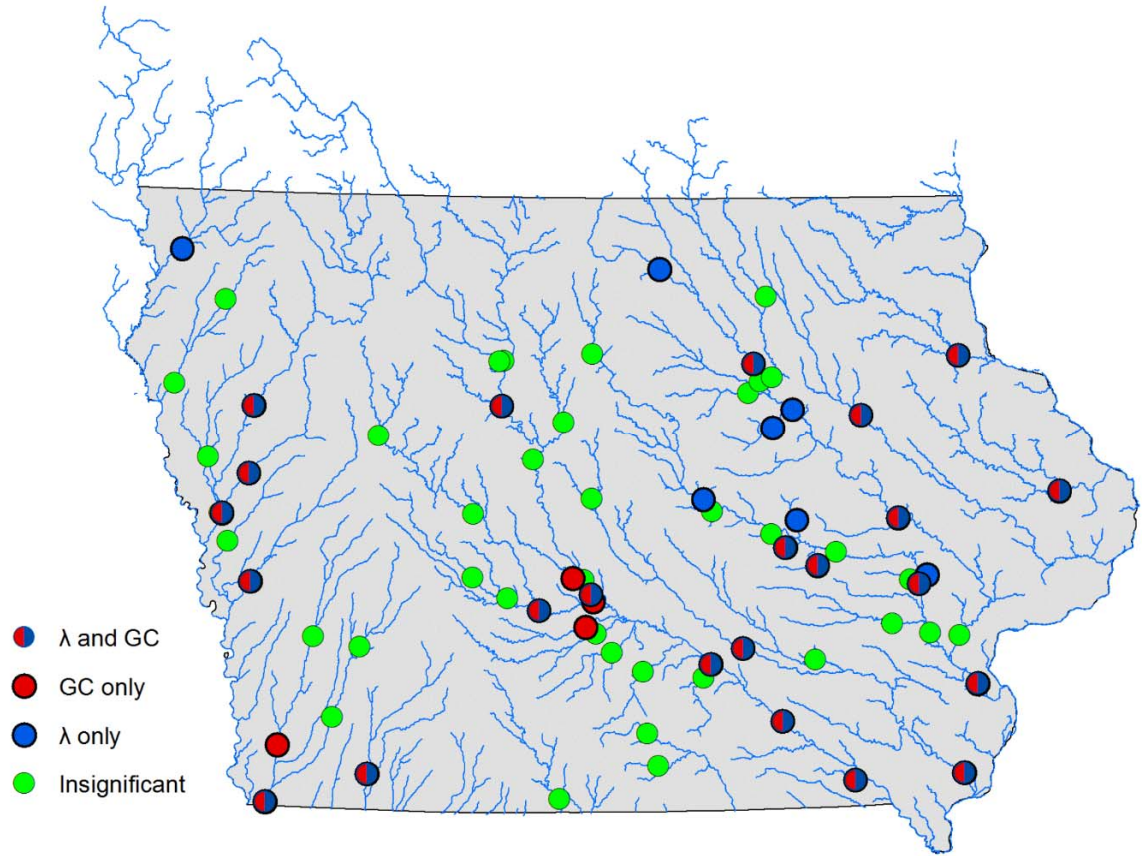

Figure 12: Illustration of USGS sites with statistically significant evolutive trends of streamflow predictability. The blue circles represent stations with evolutive trend in terms of $\lambda$ only, red circles represent stations with evolutive trend in terms of $GC$ only, circles with partial red and blue colors represent stations with evolutive trend in terms of both $\lambda$ and $GC$, while light green circles do not show any trend.

## 4 Summary and conclusions

We explored streamflow predictability across scales using seventy-one stations in the State of Iowa. Insights on the predictability of streamflow process were provided through the distinction between underlying stochastic and chaotic processes responsible for generating them, by estimating the characteristic exponent, $\lambda$ from degree distributions and the global clustering coefficient, $GC$ obtained from HVG derived complex networks. Our study answers some key questions set forth at the beginning of this paper pertaining to fundamentals of predictability of streamflow process from hydrologic standpoint.

- We showed that determining the predictability of streamflow process lies in the distinction between chaotic or stochastic processes. Our results based on HVG application to streamflow at daily time scale demonstrates that streamflow dynamics is a correlated stochastic process. The presence of correlated structure in streamflow dynamics indicates the potential for strong predictability of streamflow.
- The normalization of streamflow shows strong effect on the overall inference on predictability. We show that $GC$ is more sensitive than $\lambda$ to the form of streamflow data. The values of $GC$ show transition of dynamics regime in a large majority of networks (stations and years). It shows that normalized streamflow time series is better suited for such analyses deeming somewhat seasonal effect inherent in streamflow process. Our results in terms of $\lambda$ and $GC$ for normalized streamflow across three time-scales viz. The 15-minute, hourly and daily show the decrease as we transition to finer time-scale. The mean and median $\lambda$ falls marginally below $\lambda_{rand}$ showing that there is a tendency to transition to chaotic dynamics from stochastic dynamics as we move to finer resolution. In other words, the predictability of streamflow process through HVG based network is sensitive to time resolution of streamflow time series. A similar behavior transpires across time resolutions for $GC$. We attribute this change in streamflow predictability across time-scales to change in the dynamics of the process itself as we average it over time.
- For hydrologic community, it is of interest to understand the predictability of streamflow process across spatial scales. We show a clear spatial scale dependence of $\lambda$ for streamflow at the daily time-scale. In other words, larger the basin scales, the longer the persistent

- behavior of streamflow and hence, the predictability of streamflow. This result has important implication for water resources management and skillful streamflow forecasting.
- Finally, the forecasting community is always interested in assessing the evolutive trend of streamflow predictability. A simple linear regression-based evolution model fit shows that thirty-one stations show statistically significant trend in terms $\lambda$ while twenty-six stations show statistically significant trend in terms of $GC$. The changes could have arisen from factors such as climate change induced activities, changing rainfall patterns, and land use patterns. The explicit attribution of the trend requires further research focusing solely on their impact to predictability of streamflow process.

As rainfall is the key agent of flooding in Iowa, it should be interesting to explore the effect of rainfall on predictability of streamflow. It is clear that at small scales, rainfall and streamflow are connected stronger than at large scales, where water transport separates the two. Water transport is in the river network where streamflow aggregation process plays an important role in shaping streamflow fluctuations (e.g., Ayalew et al., 2014a, 2014b, 2015).

We neglected the effects of streamflow measurement errors as the USGS data are considered to be accurate within 5% (e.g., Ghimire and Krajewski, 2019). For our analysis, we expect the measurement uncertainty to have minimal impact on the overall inference. We performed a fundamental study on predictability of streamflow process through HVG based complex networks. We believe, our findings provide hydrologic context of interpreting underlying dynamics of streamflow process. Our analysis captures a wide range of spatial scales; hence results are deemed adequate in representing hydrologic processes across scales. Though nonlinearity is known to exist in the hydrologic process, our study does not explicitly explore nonlinearity from streamflow signals using HVG based approach. Therefore, exploiting this aspect of streamflow process and impact on predictability and incorporating them in our hydrologic modeling strategy is open to further research.

Though we implemented our analysis to Iowa, we believe that the streamflow process will demonstrate similar predictability across scales in regions with similar landscapes and climatology such as Upper Mississippi and Ohio River basins (Schilling et al., 2015) when derived from HVG based complex networks. Given that the hydro-climatologic conditions, landscapes, and base flow conditions are quite similar, the inference on streamflow dynamics across spatio-temporal scales is expected to be similar. Though we did not report it in this paper, one could separately perform the effect of time resolution (time-scale) on standard chaotic maps, which could validate our results in standalone approach.

## 5 Conflict of Interest

The authors declare that the research was conducted in the absence of any commercial or financial relationships that could be construed as a potential conflict of interest.

## 6 Author Contributions

All authors listed have made a substantial, direct and intellectual contribution to the work, and approved it for publication.

## 7 Funding

The authors received financial support for this work from the Iowa Flood Center at the University of Iowa. The third author received the additional support from the Rose & Joseph Summers endowment.

# 8 Acknowledgements

The authors acknowledge the resources help obtained from the IIHR Hydroscience and Engineering. We acknowledge the computational support from the High Performance Computing (HPC) group at the University of Iowa.